\documentclass[11pt]{article}

\usepackage{graphicx}

\usepackage[dvipdf]{epsfig}

\newcommand{\n}{\label}

\newcommand{\ben}{\begin{eqnarray}}

\newcommand{\een}{\end{eqnarray}}

\usepackage[dvipdf]{epsfig}

\usepackage{color}

\begin{document}

\title{Cosmological models with interacting components and mass-varying neutrinos}

\author{Lucas G. Collodel\footnote{lgcollodel@gmail.com}$\;$ and Gilberto M. Kremer\footnote{kremer@fisica.ufpr.br}\\
Departamento de F\'{\i}sica, Universidade Federal do Paran\'a\\ Caixa Postal 19044, 81531-990 Curitiba, Brazil}

 \maketitle

\begin{abstract}
A model for a homogeneous and isotropic spatially flat Universe, composed of baryons, radiation, neutrinos, dark matter and dark energy is analyzed. We infer that dark energy (considered to behave as a scalar field) interacts with dark matter (either by the Wetterich model \cite{Q}, or by the Anderson and Carroll model \cite{R}) and with neutrinos by a model proposed by  Brookfield et al. \cite{P}. The latter is understood to have a mass-varying behavior. We show that for a very-softly varying field, both interacting models for dark matter give the same results. The models reproduce the expected red-shift performances of the present behavior of the Universe.
\end{abstract}


\section{Introduction}

Astronomical data observation, such as measurements of type-IA supernovae \cite{A,B,C,D} indicates that our Universe is ongoing an accelerated expansion period, suggesting the existence of a different component, with the strange propriety of negative pressure, whose energy density is now predominant over others. This entity was named dark energy, and its nature still remains a mystery. Besides, the measurement of rotation curves of spiral galaxies \cite{E} is not in agreement with the luminous matter within it, proposing a new kind of matter, which does not radiate, and interacts with common components only gravitationally, named dark matter. Since those discoveries, researches have been made concerning these constituents behavior. Dark energy may be treated as the cosmological constant \cite{F}, but may also have dynamical proprieties, inciting studies with several kinds of scalar fields. {Some earlier studies considered the interaction of a scalar field with any kind of particles \cite{U}, suggesting the further treatment of particular cases.}
Furthermore, some puzzling questions arise from observation, such as the fine-tuning and the coincidence problem \cite{G,H,I,J,K}, giving rise to a whole new investigation of the components of  our Universe, in which dark energy shall interact with both dark matter and neutrino \cite{Q,R,P,L,M,N,O,S,T}.  The latter has the great advantage of experimental study, even though neutrino interaction is quite weak, the abundance of this particle allows not only detection, but opens road to investigation of its oscillating mass evolution, permitting one to develop more precise theoretical models, defining bound limits for its mass. In this paper, we work with the spatially flat Friedmann-Robertson-Walker metric, and analyze the energy densities evolution under two different dark energy-dark matter coupling, the Wetterich interaction \cite{Q,BK1} and the Anderson and Carrol interaction model \cite{R,BK2}, which we find out to be equivalent. Neutrinos are considered to present a mass-varying behavior and to be coupled to dark energy by a model proposed by Brookfield et al. \cite{P}. Natural units were rescaled ($8\pi G=c=1$).

\section{Field Equations}

We consider a Universe composed by the following fields: baryons $(b)$, radiation $(r)$, dark energy $(\varphi)$, dark matter $(dm)$ and neutrinos $(\nu)$. The baryons and radiation are regarded as non-interaction fields, the dark energy is interpreted to behave as a scalar field which interacts with the dark matter and the neutrinos are treated as a mass-varying component due to a coupling to the dark energy field.

 In a conformal time the Friedmann-Robertson-Walker metric for an isotropic, homogeneous and spatially flat Universe is given by
\begin{equation}
 ds^2= a(\tau)^2\left[d\tau^2 -\delta_{ij}dx^idx^j \right],
\end{equation}
where $a(\tau)$ represents the cosmic scale factor.

 The Friedmann and acceleration equations  which follows from Einstein field equations in conformal time read
\begin{equation}\n{2}
\left(\frac{\dot{a}}{a}\right)^2=\frac{1}{3}\rho a^2, \qquad \frac{\ddot{a}}{a}=\frac{1}{6}(\rho-3p)a^2,
\end{equation}
respectively. The dots denote the differentiation with respect to the conformal time, while $\rho= \rho_{b}+\rho_{r}+\rho_{dm}+\rho_{\nu}+\rho_{\varphi}$ and $p=p_{b}+p_{r}+p_{dm}+p_{\nu}+p_{\varphi}$ are the total energy density and pressure of the sources, respectively.

From the  conservation of the energy-momentum tensor it follows the evolution equation for the total energy density:
\begin{equation}\n{3}
\dot{\rho}+3\frac{\dot{a}}{a}(\rho+p)=0.
\end{equation}

 The baryons are treated like dust ($p_b=0$) and the barotropic equation of state for the radiation field is $p_r=\rho_r/3$. Once both of them have no interaction with other components, the evolution equations for their energy densities are:
\begin{equation}\n{4}
\dot{\rho}_b+3\frac{\dot{a}}{a}\rho_b=0, \qquad \dot{\rho}_r+4\frac{\dot{a}}{a}\rho_r=0,
\end{equation}
 respectively.

Since the dark energy is modeled as a scalar field $\varphi$ its  energy density and pressure are given by
\begin{equation}\n{5}
\rho_{\varphi}=\frac{1}{2a^2}\dot{\varphi}^2+V(\varphi), \qquad p_{\varphi}=\frac{1}{2a^2}\dot{\varphi}^2-V(\varphi),
\end{equation}
where $V(\varphi)$ denotes the  potential of the scalar field.

 In this work we shall analyze two types of interactions between dark matter and dark energy. The first one is the Wetterich  model \cite{Q} which for a pressureless ($p_{dm}=0$) interacting dark matter has the following evolution equation
\begin{equation}\n{6}
\dot{\rho}_{dm}+3\frac{\dot{a}}{a}\rho_{dm}=-\beta\rho_{dm}\dot{\varphi}.
\end{equation}

The other refers to the Anderson and Carroll model \cite{R} and in this case the evolution equation for the dark matter reads
\begin{equation}\n{7}
\dot{\rho}_{dm}+3\frac{\dot{a}}{a}\rho_{dm}=\beta\rho_{dm}\frac{\dot{\varphi}}{\varphi}.
\end{equation}
{In both equations above, $\beta$ stands for the coupling constant between dark matter and dark energy, which may be interpreted as the fundamental scalar charge $q$ of the particles \cite{BB}.}

The neutrinos are understood to be massless and relativistic particles in the past, but with the coupling to the dark energy, they have acquired mass and became non-relativistic, having an oscillating mass behavior at low red-shifts. The evolution equation for its energy density according to Brookfield et al. \cite{P} is
\begin{equation}\n{8}
\dot{\rho}_{\nu}+3\frac{\dot{a}}{a}(\rho_{\nu}+p_{\nu})=\frac{d\ln{m_{\nu}}}{d \varphi}\dot{\varphi}(\rho_{\nu}-3p_{\nu}),
\end{equation}
where $m_\nu$ is the neutrino mass which is considered a function of the scalar field $\varphi$. Here, we follow \cite{P} and write the neutrino mass as $m_{\nu}=M_0  e^{\alpha   \varphi}$,
where $M_0$ and $\alpha$ denote coupling constants.

From the evolution equation for the total energy density (\ref{3}) and by taking into account (\ref{4}), (\ref{8}) and (\ref{6}) or (\ref{7}), we may obtain two evolution equations for the energy density of the scalar field, one referring to the Wetterich model and the other to the Anderson and Carrol model. Furthermore, from the resulting equations it is possible to obtain evolution equations for the scalar field by considering the representations for its energy density (\ref{5})$_1$ and for its pressure (\ref{5})$_2$. These equations reads
\begin{equation}\n{9}
\ddot{\varphi}+2\dot{\varphi}\frac{\dot{a}}{a}+a^2\left[\frac{dV}{d\varphi}+\frac{d\ln m_{\nu}}{d \varphi}(\rho_{\nu}-3p_{\nu})-\beta\rho_{dm}\right]=0,
\end{equation}
\begin{equation}\n{10}
\ddot{\varphi}+2\dot{\varphi}\frac{\dot{a}}{a}+a^2\left[\frac{dV}{d\varphi}+\frac{d\ln m_{\nu}}{d \varphi}(\rho_{\nu}-3p_{\nu})+\beta\frac{\rho_{dm}}{\varphi}\right]=0,
\end{equation}
where the first one refers to the Wetterich model, while the second one to the Anderson and Carrol model.

From now on we shall use  the red-shift $z$ as variable instead of the conformal time $\tau$. Since the red-shift is represented in terms of the cosmic scale factor by $a=1/(1+z)$, we have the following relationships:
\begin{equation}\n{11}
\frac{d}{d\tau}=-\sqrt{\frac{\rho}{3}}\frac{d}{dz},\qquad
\frac{d^2}{d\tau^2}= \frac{\rho+p}{2}\frac{1}{1+z}\frac{d}{dz} +\frac{\rho}{3}\frac{d^2}{dz^2}.
\end{equation}

The differential equations for baryons (\ref{4})$_1$ and radiation (\ref{4})$_2$ can be integrated and this yields the well known relationships:
\begin{equation}\n{12}
\rho_b(z)=\rho_b^0(1+z)^3, \qquad \rho_r(z)=\rho_r^0(1+z)^4,
\end{equation}
 where the index 0 stands for the present value of the variable.

 For the energy density of the dark matter, we obtain through the integration of (\ref{9}) and (\ref{10})
\ben\n{13}
\rho_{dm}(z)&=&\rho_{dm}^0(1+z)^3 e^{-\beta\left[\varphi(z)-\varphi(0)\right]},
\\\n{14}
\rho_{dm}(z)&=&\rho_{dm}^0(1+z)^3\left( \frac{\varphi(z)}{\varphi(0)} \right)^\beta,
\een
for the Wetterich and Anderson and Carroll models, respectively.

In terms of the red-shift the differential equation for the energy density of the neutrinos (\ref{8}) becomes:
\begin{equation}\n{15}
\rho'_{\nu}=3\frac{\rho_{\nu}+p_{\nu}}{1+z}+\frac{d\ln m_{\nu}}{dz}(\rho_{\nu}-3p_{\nu}),
\end{equation}
where the prime denotes differentiation with respect to the red-shift ($z$).

Furthermore, the differential equations for the scalar field in terms of the red-shift for the Wetterich and Anderson and Carroll models -- which follow from (\ref{9}) and (\ref{10}) -- read
\begin{eqnarray}\n{16}
\frac{\rho}{3}\varphi''+\frac{3p-\rho}{6(1+z)}\varphi'
+\frac{1}{(1+z)^2}\bigg[ \frac{dV}{d\varphi}
+\frac{d\ln m_{\nu}}{d \varphi}(\rho_{\nu}-3p_{\nu})-\beta\rho_{dm}\bigg]=0,
\\\n{17}
\frac{\rho}{3}\varphi''+\frac{3p-\rho}{6(1+z)}\varphi'
+\frac{1}{(1+z)^2}\bigg [\frac{dV}{d\varphi}
+\frac{d\ln m_{\nu}}{d \varphi}(\rho_{\nu}-3p_{\nu})+\beta\frac{\rho_{dm}}{\varphi}\bigg]=0,
\end{eqnarray}
respectively.

Due to the fact that  the energy densities of the baryons, radiation and dark matter are determined by the expressions (\ref{12}) -- (\ref{14}), the equations (\ref{15}) -- (\ref{17}) compose our system of coupled differential equations for $\rho_\nu$ and $\varphi$ we are willing to solve.

Once the neutrino mass was supposed to have the form $m_{\nu}=M_0  e^{\alpha   \varphi}$,  we have to specify the potential of the scalar field $V(\varphi)$ and the equation of state of the neutrinos $p_\nu=p_\nu(\rho_\nu)$ in order to solve the system of coupled differential equations.

Since authors began to study neutrinos coupled to dark energy, many kinds of pressure have been attributed to it. Some of them suggest that the pressure would be proportional to their density through a barotropic equation of state $p_{\nu}=w_{\nu}\rho_{\nu}$, where $w_{\nu}$ is a constant. Note that if $w_{\nu}=1/3$, the coupling vanishes making neutrinos non-interacting particles. Recently, the neutrino varying-mass became a strong aim of study among researchers and this emerges the need of investigating varying-pressures, so that we could better explain the coupling mechanism. Here, we introduce two possible  equations of state:
\begin{equation}\n{18}
p_{\nu}=\frac{z}{3(1+z)}\rho_{\nu}, \qquad p_{\nu}=\frac{1-e^{-z}}{3}\rho_{\nu},
\end{equation}
in accordance with what we postulated before. We should notice that as the red-shift tends to high values, these equations of state turn out to be equivalent to the case of radiation ($p_\nu=\rho_\nu/3$) during the period of non-coupling with the dark energy, while neutrinos were still relativistic. In the recent era the neutrinos are  pressureless ($p_\nu=0$) and they are treated as dust particles. Note that the second equation of state tends more rapidly to  ($p_\nu=\rho_\nu/3$) for large values of the red-shift $z$.
{Although we have proposed the two above equations of state for the description of the past and present behavior of the Universe, one may observe that the extrapolation to negative values of the redshift will lead to negative values of the neutrino pressure field and even the first equation will diverge at $z=-1$. Hence, the two equations of state proposed here cannot be extrapolated to negative values of the redshift.}

The potential adopted in this work is the "Mexican hat" potential:
\begin{equation}
V(\varphi)=\lambda \left[\varphi(z)^2-\varphi_0^2 \right]^2-\lambda\varphi_0^4,
\end{equation}
where $\lambda$ and $\varphi_0$ denote two constants.

\section{Solutions}

The numerical solutions of the system of equations (\ref{15}) and (\ref{16}) and of (\ref{15}) and (\ref{17}) -- according to the Wetterich and Anderson and Carrol models -- were found  by considering the equations of state (\ref{18}) and the specification of the coupling and potential parameters together with the initial conditions.

First we introduced the density parameters $\Omega_i=\rho_i/\rho$ and their values at $z=0$: $\Omega_b(0)=0.05$, $\Omega_r(0)=5\times10^{-5}$ and $\Omega_{dm}(0)=0.23$.
Next the initial conditions for $\varphi$, $\varphi'$ and $\rho_\nu$ were specified at $z=3100$, namely, $\varphi(3100)=\varphi_0=2.69$ and $\varphi'(3100)\approx0$ and $\Omega_\nu(3100)\approx0.68\,\Omega_r(3100)$ (see e.g.\cite{Liddle}).
{Furthermore,  as in any phenomenological theory, the investigation of the interactions in the dark sector  requires the knowledge of some  values for the coupling constants, in order to reproduce the observed behavior of the present Universe. Here we have the freedom to choose the potential constant and the dark matter coupling parameter. The potential constant  was chosen to be $\lambda=-0.015$, and the dark matter coupling parameter: $\beta=-0.01$ in the Wetterich model and $\beta=0.03$ for the Anderson and Carroll interaction.}

The scalar field, for both models, varies pretty slowly during the expansion of the Universe (see Fig. 1), which means that the energy density and the pressure of the dark energy have always been dominated by its potential and, except for its tiny variation, it behaves like a cosmological constant. For this reason, the coupling between neutrino and dark energy must be strong. Indeed, such couplings are very unusual and not easy-to-swallow, once free cosmological parameters are often found in a range $[-1,1]$, suggesting a re-scale of the scalar field: $\tilde{\varphi}=10 \varphi + K$ where $K=-23.83$ is a constant chosen to keep the initial conditions. The determination of the coupling parameter $\alpha$  was found through the variation of its value and by constraining that the neutrinos mass shall present an oscillating evolution, as well as assuring that the dark energy density parameter must
not grow at large red-shifts. Once we did this, our coupling parameter to neutrinos is $\alpha=10$, still a high value, but acceptable considering the field behavior, and the investigation of new forces.

\begin{figure}
\begin{center}
\includegraphics[scale=0.3]{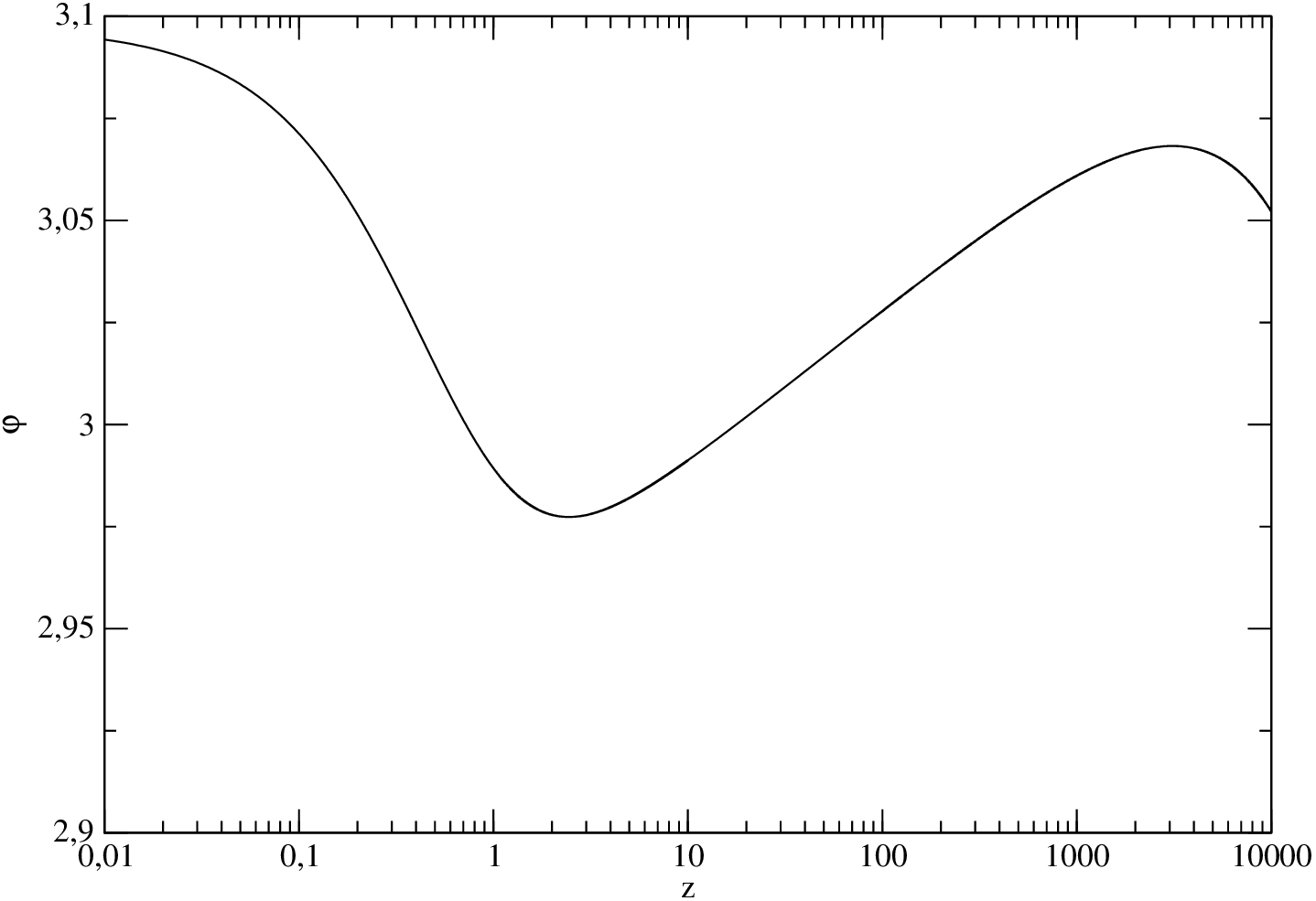}
\caption{Scalar field vs. red-shift in logarithmic scale for $\alpha=10$ and $M_0=10^{-14}$.}\vskip0.5cm
\end{center}
\end{figure}

{The scalar field varying very softly during all times, exactly the same way for both models, and not presenting singularities, resulted in the pretty same curves for all four models (considering now the two barotropic equations for neutrinos). {Nevertheless, we notice from Figure 1 an oscillating behavior in the scalar field, which increases its value as the redshift approaches zero. Thus we may already expect a similar behavior for neutrinos mass.} We can see in Figures 2 and  3, the evolutions of the density parameters, which occurred as expected: at high red-shifts, dark energy is neglected, the radiation and baryons fall with $a^{-4}$ and $a^{-3}$ respectively. We note from Fig. 3 that the density parameter of the dark matter presents a very similar behavior as the baryons. This fact can be understood as follows: besides the energy transfer of the dark components, dark matter also falls with $a^{-3}$, once the amount of dark energy is so small for the period shown in this figure. Furthermore, as can be observed from Fig. 2 the evolution of the dark energy  density does not show up in Fig. 3, because it decreases very rapidly with the red-shift.  In this model the transition between two important epochs of our Universe, the radiation-matter equality ($\Omega_{dm}+\Omega_b=\Omega_r+\Omega_\nu$) takes place at $z\sim 3400$. The neutrino showed a moderate energy density in past, decreasing very softly until recent times so that $\Omega_\nu(0)\approx1.4\times 10^{-4}$. For small red-shift values, the dark energy becomes dominant very fast, reaching $\Omega_\varphi(0)\approx0.71$ and making the Universe to  expand acceleratedly, as we can see in Fig. 4, where we plotted the deceleration parameter, given by:
$q=\frac{1}{2}+\frac{3p}{2\rho}$. For $z=0$ the deceleration parameter has the value $q(0)\approx-0.53$, while  the transition decelerated-accelerated occurs  at $z_t\approx0.74$, both values are of the same order of magnitude of the  ones given in the literature: $q(0)=-0.46\pm0.13$ (see~\cite{Vir}) and
$z_t=0.74\pm0.18$ (see~\cite{riess}).

\begin{figure}
\begin{center}
\includegraphics[scale=0.3]{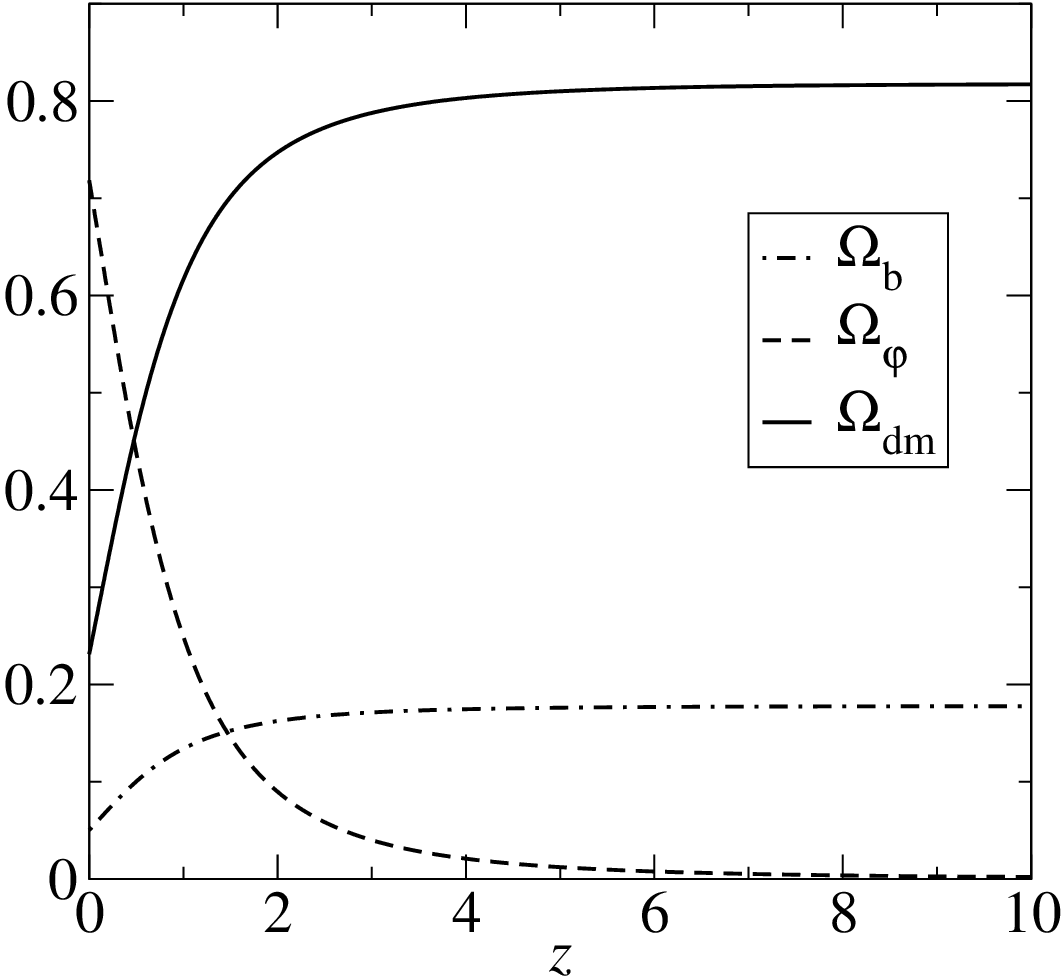}
\caption{Density parameters vs. red-shift.}
\end{center}
\end{figure}

\begin{figure}
\begin{center}
\includegraphics[scale=0.3]{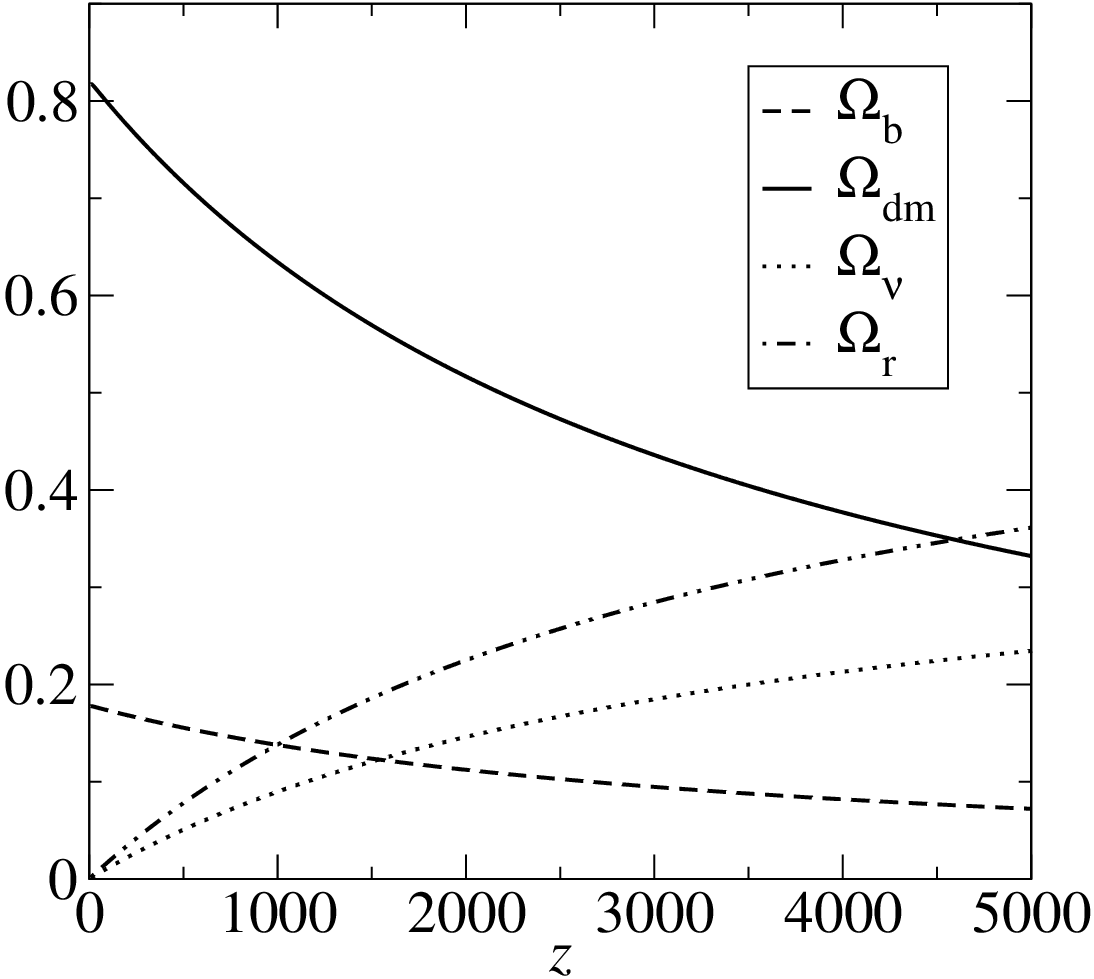}
\caption{Evolution of density parameters in large scale.}
\end{center}
\end{figure}

\begin{figure}
\begin{center}
\includegraphics[scale=0.3]{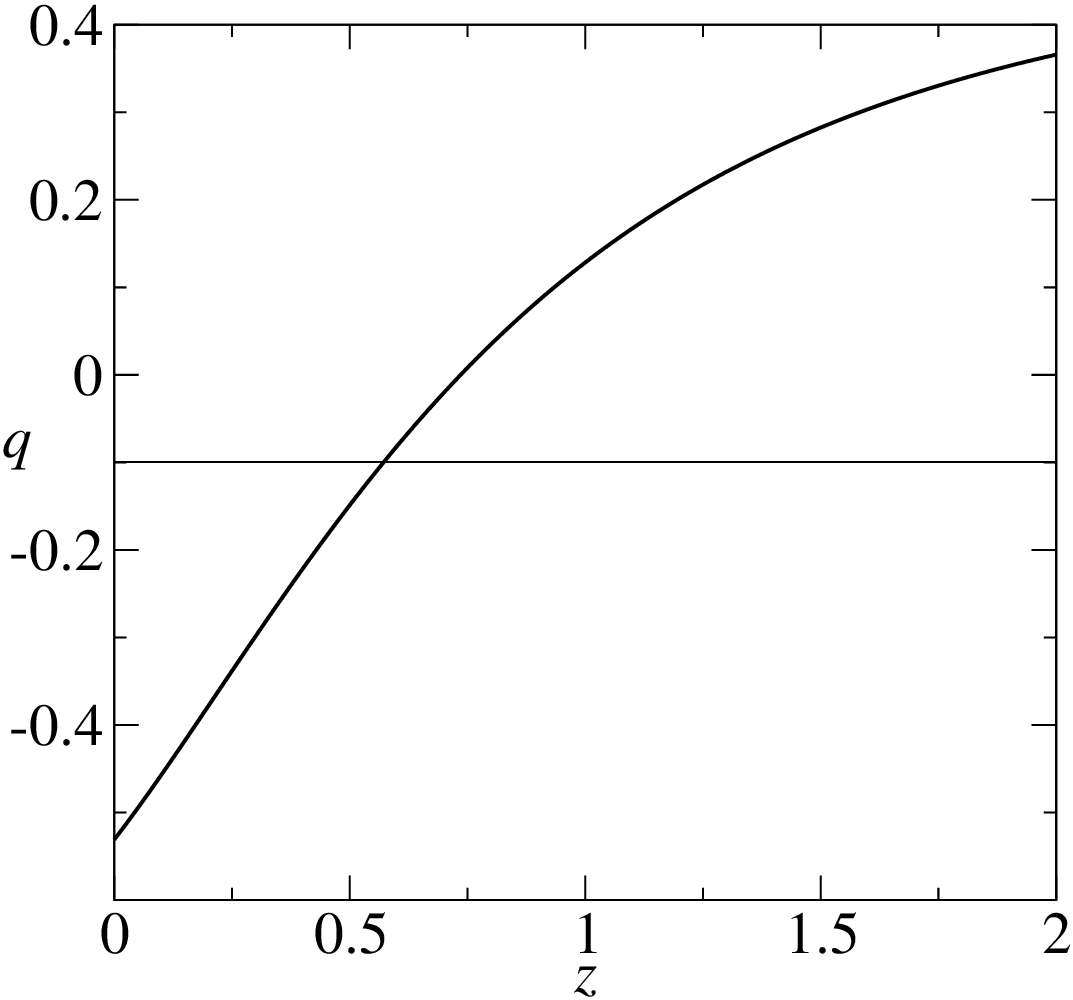}
\caption{Deceleration parameter vs. red-shift.}
\end{center}
\end{figure}

\begin{figure}
\begin{center}
\includegraphics[scale=0.3]{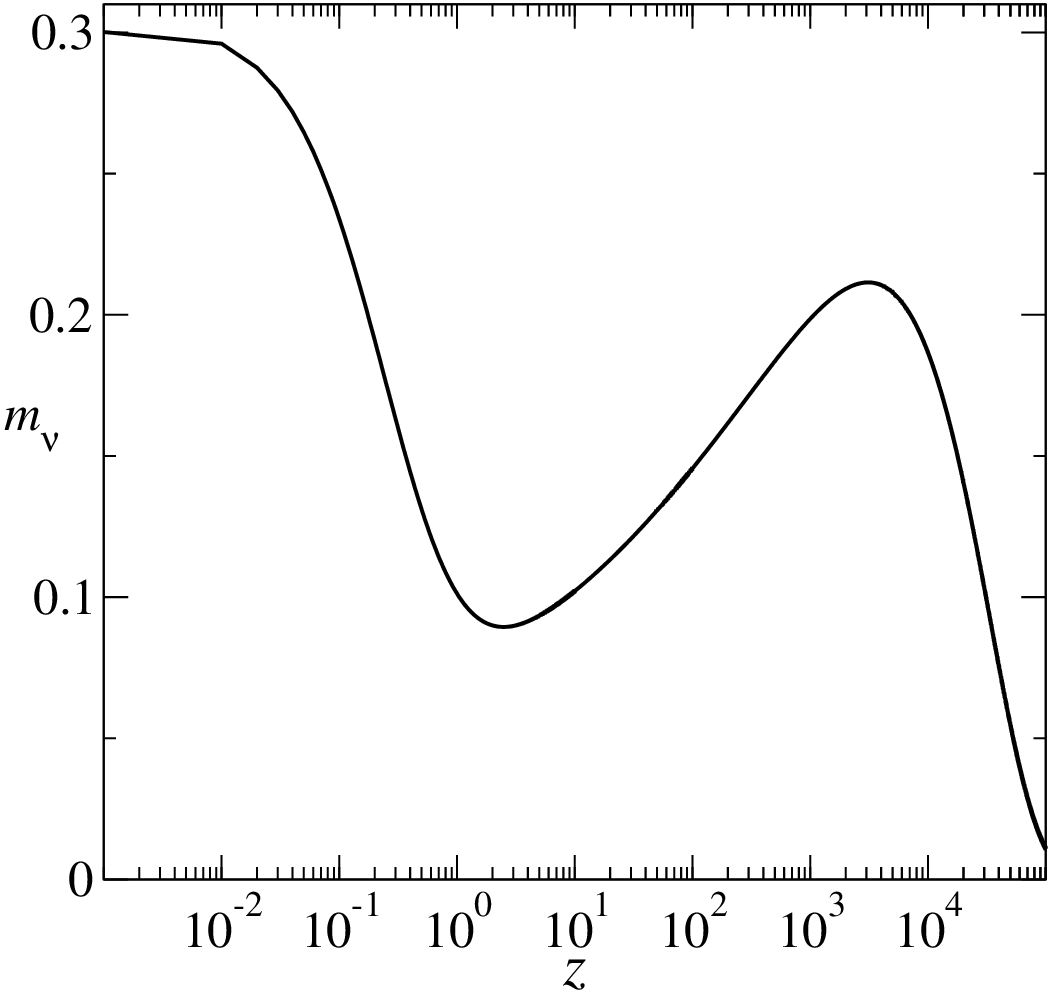}
\caption{Evolution of neutrino mass vs. red-shift in logarithmic scale.}
\end{center}
\end{figure}

The mass evolution is plotted in Fig. 5, the constant $M_0$ in the expression $m_{\nu}=M_0  e^{\alpha   \varphi}$ plays an essential role as a free parameter, and here we chose it to be $M_0=10^{-14}$. It is clear that it does not change the shape of the curve (the neutrinos mass behavior does not depend on it) but instead, changing its value translates the curve upwards or downwards. We see from this figure that for recent times,the  neutrino mass would be about $0.31 eV$. If we go back in time, as the red-shift increases, so does the neutrino mass, which evolves until $z\sim3000$ reaching $m_{\nu}=0.21 eV$ and then begins to decrease again until zero at large red-shifts.}

\section{Conclusions}

{Although we chose to work with the "Mexican hat" potential, other kinds like the Yukawa potential have been investigated, giving the same soft variation of the field. For this reason, for any red-shift value, $\varphi'$ assumes no predominant position, making the field pressure and energy density dominated by its potential. Furthermore, the almost constant behavior of the field, makes both interacting models differ only by a constant, which is contoured by attributing correct values to the coupling parameters, resulting in equivalent models. As mentioned before, it suggests a very strong dark energy-neutrino coupling, if we are really attained to neutrino mass-varying property, leading to the assumption that the deep investigation of the latter will improve our understanding of the first.

\end{document}